\RequirePackage{etoolbox}
\csdef{input@path}{{style/}{graphics/}}
\documentclass[aoas,MSNbibl,nameyear,dvips]{arximspdf}
\makeatletter
   \@ifpackageloaded{graphicx}{}{\usepackage{graphicx}}
\makeatother
\usepackage{dcolumn}
\usepackage{graphicx}

\doi{10.1214/14-AOAS801}
\volume{9}
\issue{1}
\pubyear{2015}
\firstpage{1}
\lastpage{26}
\docsubty{FLA}

\makeatletter
\newcolumntype{d}[1]{D{.}{.}{#1}}
\newcommand{\rrvert}{\vert}
\newcommand{\rrVert}{\Vert}
\newcommand{\llvert}{\vert}
\newcommand{\llVert}{\Vert}
\newcommand{\mcal}[1]{\mathcal{#1}}
\newcommand{\mbb}[1]{\mathbb{#1}}
\def\log{\operatorname{log}}
\makeatother

\begin{document}
\begin{frontmatter}

\title{Bayesian binomial mixture models for estimating abundance in
ecological monitoring studies\thanksref{T1}}
\runtitle{Bayesian binomial mixture models}

\begin{aug}
\author[A]{\fnms{Guohui}~\snm{Wu}\thanksref{M1}},
\author[B]{\fnms{Scott H.}~\snm{Holan}\corref{}\thanksref{M2}\ead[label=e1]{holans@missouri.edu}},
\author[C]{\fnms{Charles H.}~\snm{Nilon}\thanksref{M2}}\\
\and
\author[B]{\fnms{Christopher K.}~\snm{Wikle}\thanksref{M2}}
\runauthor{Wu, Holan, Nilon and Wikle}
\affiliation{SAS Institute Inc.\thanksmark{M1} and University of
Missouri\thanksmark{M2}}
\address[A]{G. Wu\\
SAS Institute Inc.\\
SAS Campus Drive\\
Cary, North Carolina 27513\\
USA}
\address[B]{S. H. Holan\\
C. K. Wikle\\
Department of Statistics\\
University of Missouri\\
146 Middlebush Hall\\
Columbia, Missouri 65211-6100\\
USA\\
\printead{e1}}
\address[C]{C. H. Nilon\\
Department of Fisheries and Wildlife Sciences\\
University of Missouri\\
Anheuser-Busch Natural Resources Building\\
Columbia, Missouri 65211-7240\\
USA}
\end{aug}
\thankstext{T1}{Supported in part by the U.S. National Science Foundation (NSF) and the U.S. Census Bureau under NSF Grant SES-1132031,
funded through the NSF-Census Research Network (NCRN) program. Additionally, this research was supported  in part by
NSF DMS-10-49093 and the NSF Long-Term Ecological Research Program (DEB-1027188).}

%
\received{\smonth{4} \syear{2014}}
%
\revised{\smonth{12} \syear{2014}}

%
\begin{abstract}
Investigation of species abundance has become a vital component of many
ecological monitoring studies. The primary objective of these studies
is to understand how specific species are distributed across the study
domain, as well as quantification of the sampling efficiency for
detecting these species. To achieve these goals, preselected locations
are sampled during scheduled visits, in which the number of species
observed at each location is recorded. This results in spatially
referenced replicated count data that are often unbalanced in structure
and exhibit overdispersion. Motivated by the Baltimore Ecosystem Study,
we propose Bayesian hierarchical binomial mixture models, including
Binomial Conway--Maxwell Poisson (Bin-CMP) mixture models, that formally
account for varying levels of spatial dispersion. Our proposed models
also allow for variable selection of model covariates and grouping of
dispersion parameters through the implementation of reversible jump
Markov chain Monte Carlo methodology. Finally, using demographic
covariates from the American Community Survey, we demonstrate the
effectiveness of our approach through estimation of abundance for the
American Robin (\textit{Turdus migratorius}) in the Baltimore
Ecosystem Study.
\end{abstract}

%

\end{frontmatter}

\setcounter{footnote}{1}

\section{Introduction}
Investigation of species abundance is a topic of\break widespread interest in
ecology. To estimate and model variation in species abundance,
predetermined survey points are visited at each sampling occasion and
the number of animals detected are recorded. This results in spatially
referenced point count data. Such a sampling protocol is easier to
implement than the traditional capture--recapture experiment [e.g., see
\citet{williams2002analysis} and the references therein], since
each animal encountered does not have to be distinctly tagged.
Nevertheless, these spatially referenced data can be utilized to
estimate the abundance of animals, for which individual tagging might
be difficult or even infeasible due to the amount of effort involved,
for example, in some avian ecology surveys. Therefore, to estimate
abundance, the development of binomial mixture models has drawn
significant attention over the past few decades [e.g., \citet
{carroll1985note,royle2004n,kery2005modeling,kery2008estimating,webster2008bayesian}].

In developing statistical models for count data, the choice of the
distribution function frequently depends on the dispersion associated
with the data. For equidispersed data (i.e., equal mean and variance),
the Poisson distribution is frequently used due to its explicit
assumption of equidispersion. However, to model overdsipersed data
(i.e., the variance is greater than the mean), the choice of
distribution functions needs to be made [e.g., see \citet
{ver2007quasi}]. Often, the negative binomial (NB) distribution
[\citet{cameron1998regression}] is employed, due to a dispersion
parameter that conveniently controls the level of overdispersion.
Alternatively, the Poisson distribution can also be used with a random
effect included to relax the restrictive assumption of equidispersion.
Although the Poisson and NB distributions have become the {de
facto} options for count data, neither of them accounts for
underdispersion (i.e., the variance is less than the mean). Admittedly,
overdispersion is more common for data arising from ecological
monitoring studies, while underdispersion is often present for rare
event data [e.g., \citet
{herbers1989community,ridout2004empirical,oh2006accident}].
Nevertheless, cases can arise in ecological monitoring studies where
the species of interest is less prevalent (due to being rare
occurrences). In principle, these situations would manifest themselves
as underdispersion.

The Conway--Maxwell Poisson (CMP) distribution [\citet
{conway1962queuing}] is an ideal candidate for modeling count data with
different types of dispersion, as it has an extra dispersion parameter
that flexibly allows for equi-, over-, and underdispersion. Moreover,
the CMP distribution is closely related to many other discrete
distributions. For example, the CMP distribution contains the Poisson
distribution as a special case and generalizes Bernoulli and geometric
distributions in the limiting cases [\citet{shmueli2005useful}].
Owing to its versatility, the CMP distribution has become increasingly
popular among many subject-matter disciplines. For example, in the
context of breeding bird surveys,
\citet{wuCMP2013} develop a Bayesian hierarchical spatio-temporal
CMP model for complex and high-dimensional count data. A unique aspect
of this research is that it allows for dynamic spatial dispersion
(i.e., the dispersion over the spatial domain evolves over time). A
comprehensive overview regarding the CMP model is provided by
\citet{sellers2011poisson} and the references therein.

Binomial mixture models have become increasingly popular for analyzing
spatial point referenced count data in the context of estimating and
modeling variation in species abundance. As a result, various models
have been developed with this application in mind. For example,
\citet{carroll1985note} consider a Binomial-Beta mixture model to
study the problem of estimating an unknown population, $N$, that
follows a discrete uniform distribution, in which efficient estimators
were obtained through the use of an integrated likelihood method. To
improve the estimator proposed by \citet{carroll1985note},
\citet{royle2004n} develops a Binomial--Poisson (Bin--Pois) mixture
model, in which $N$ is considered to be an independent random variable
from a Poisson distribution. Subsequently, \citet
{royle2006hierarchical} propose a more general hierarchical modeling
framework with the goal of addressing animal abundance in the case of
imperfect detection, wherein the variation associated with the observed
data was partitioned into that of abundance and that of detectability.
In the context of avian ecology studies, \citet{kery2005modeling}
and \citet{kery2008estimating} apply the Bin--Pois models to the
estimation of bird abundance. \citet{webster2008bayesian} propose
a Bin--Pois model, in which a conditional autoregressive (CAR) model was
used to address spatial dependence found in the bird density.
\citet{wenger2008estimating} develop zero-inflated Bin--Pois and
zero-inflated Binomial--negative binomial (Bin--NB) models for the
estimation of species abundance. \citet{kery2010hierarchical}
develop a Bin--Pois model with a site-specific random effect to allow
for overdispersion and, thus, the equidispersion assumption of the
Poisson distribution is relaxed. \citet{graves2011linking} apply
the Bin--Pois model to estimate abundance for a grizzly bear population
using multiple detection methods, in which covariates are introduced to
explain variation in both the detection and intensity process. Under
the frequentist framework, \citet{dail2011models} propose a
general Bin--Pois model to allow for a formal statistical test regarding
the assumption of population closure. However, none of the
aforementioned models simultaneously allows for data with different
levels of dispersion (over- and underdispersion) and Bayesian model
selection (e.g., using the Conway--Maxwell Poisson distribution and
reversible jump Markov chain Monte Carlo).

Some experiments in ecological studies can be viewed as a robust design
[e.g., see \citet{pollock1982capture}], that is, there are
secondary, and possibly subsequent, sampling periods nested within each
primary sampling occasion. For example, the American Robin (\textit
{Turdus migratorius}) data we consider from the Baltimore Ecosystem Study (BES) falls into this category. This nested sampling design
contains the design with one primary sampling occasion as a special
case. Motivated by American Robin data from BES (Section~\ref
{secApp}), we develop a Binomial Conway--Maxwell Poisson (Bin-CMP)
mixture model that accommodates both overdispersed and underdispersed
data under a nested/unbalanced data structure. The Bin-CMP models we
propose are cast in a general Bayesian hierarchical binomial \mbox{mixture}
model framework that can accommodate mixtures using distributions other
than the CMP.

Compared with the existing models in the literature, our contribution
can be seen as follows. First, we develop a flexible class of binomial
mixture models to account for replicated count data with different
types of dispersion, which is achieved by choosing a suitable model for
the abundance parameter (e.g., using the CMP distribution). In the case
of overdispersed data, our methodology is advantageous from an
estimation perspective when compared to the general\vadjust{\goodbreak} modeling strategy
that includes a random effect to account for extra dispersion [e.g.,
see \citet{kery2010hierarchical}], as our model has a fewer
number of parameters to be estimated. Although each parameter may be
more computationally expensive, compared to the strategy of including a
random effect, this computational burden can be alleviated through the
use of a lower level programing language and parallel computation. More
importantly, our model provides an explicit quantification of
dispersion and can also be used in the context of underdispersed data.
Additionally, the models we consider can flexibly account for spatial
dependence in species abundance by adding a low-rank spatial component
to the model for the intensity process. In contrast to the CAR models
used by \citet{webster2008bayesian}, our methodology does not
require us to define a neighborhood structure for the point count data,
which can be difficult in many cases. In the setting of our motivating
example, where the bird counts themselves are modeled at the point
level rather than on areal units, a geostatistical approach may be more
appropriate. Further, through reversible jump Markov chain Monte Carlo
(RJMCMC), we introduce automated variable selection for covariates and
grouping of dispersion parameters into the binomial mixture modeling
framework and, to the best of our knowledge, our approach constitutes
the first successful RJMCMC implemented on the CMP dispersion
parameters. Last, the variable selection allows us to identify
important predictors related to high detectability and abundance for a
given species of interest.

This paper is organized as follows. Section~\ref{secdataprelim}
introduces our motivating data from the BES and provides preliminary
background information on the CMP distribution. Section~\ref
{secModeldev} describes our proposed Bayesian hierarchical binomial
mixture models, including the Bin-CMP model. Section~\ref{secModelsel}
provides relevant information on Bayesian variable selection and
grouping using RJMCMC. Simulated examples are presented in Section~\ref
{secSim}, illustrating the effectiveness of our modeling approach.
Section~\ref{secApp} contains an analysis of our motivating data,
estimating abundance of the American Robin from the BES, and
demonstrates the utility of our methodology. Discussion is provided in
Section~\ref{secDiscu}. For convenience of exposition, specific
details surrounding our Markov chain Monte Carlo (MCMC) algorithm and
full conditional distributions are left to a supplemental article
[\citet{wuSupp2015}].

\section{Data and preliminary background}\label{secdataprelim}
\subsection{Baltimore Ecosystem Study survey data}

\begin{figure}

\includegraphics{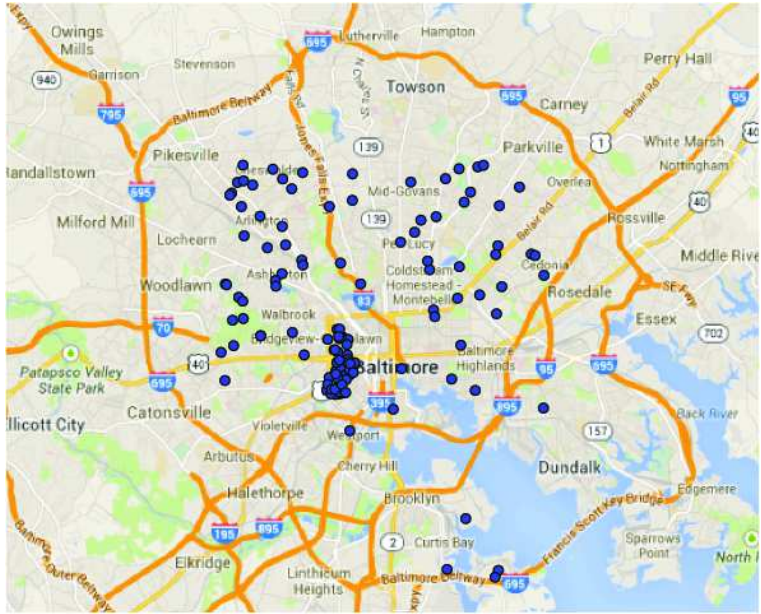}

\caption{Plot of 131 bird census points for
American Robin in the City of Baltimore, Maryland (using R package
``RgoogleMaps''). The solid circles are bird census points.}\label{figeustobsdata}
\end{figure}

As a long-term ecological monitoring study, the BES considers the City
of Baltimore, Maryland as a study area, with the objective of
understanding how the City of Baltimore evolves as an ecosystem over
time [\citet{pickett2011urban}]. Collected as a part of the BES,
the American Robin (\textit{Turdus migratorius}) data we consider
constitutes spatially replicated point count data on 132 bird census
points in the City of Baltimore, which are randomly selected from a set
of urban forest effect (UFORE or I-Tree Eco) model points (Section~\ref
{secApp}). Considered as the most widespread North American thrush, the
American Robin has become common in many North American cities
[\citet{sallabanks1999american}]. Despite its abundance,
conservation measures, which are enforced by the Migratory Bird Treaty
Act of 2004,
have been taken to protect the American Robin throughout its
geographical range in the United States.
Although BES data have been collected across bird survey points since
2005, as an illustration, we consider a subset of data over five years
from 2005 to 2009, due to incomplete data in later years. In each year,
three surveys were scheduled for each of the survey points throughout
May and August, each of which consisted of a five minute survey
conducted between 5 am and 10 am on days without rain. During each
survey, the recorded count represents the combination of birds that
were seen, heard, or flew over each survey point. In the current
context, the secondary sampling period consists of the five minute
daily survey, while the primary sampling periods are the time frames
determined by the dates on which three daily surveys are conducted. As
a result, the nested sampling design provides a maximum of 15 spatially
referenced counts for each bird census point. Despite the fact that
several species are available in the BES, as an illustration, we
consider American Robin counts in our analysis, due to their higher
abundance relative to other species. Among the 132 bird census points,
131 of them have American Robin detections (Figure~\ref{figeustobsdata}).

%

\subsection{The Conway--Maxwell Poisson distribution}

Let $X$ denote a CMP distributed random variable, that is, $X \sim
\operatorname{CMP}(\lambda,\nu)$, where $\lambda>0$ and $\nu\geq0$ are
the CMP intensity and dispersion parameters, respectively. The
probability mass function (pmf) of $X$ is given by
%
\begin{equation}\label{eqcmppdf}
P(X=x)=\frac{\lambda^{x}}{(x!)^\nu}\frac{1}{Z(\lambda,\nu)},\qquad
x=0,1,2,\ldots,
\end{equation}
where
%
\begin{equation}
Z(\lambda,\nu)=\sum_{j=0}^{\infty}{
\frac{\lambda^{j}}{(j!)^\nu}} \label{eqcmpzfun}
\end{equation}
is a normalizing constant (often referred to as the ``$Z$-function'').
With the additional parameter $\nu$, the CMP distribution conveniently
accommodates equidispersion, overdispersion, and underdispersion.
Specifically, $\nu=1$ corresponds to the Poisson distribution, whereas
$\nu<1$ and $\nu>1$ represent overdispersion and underdispersion,
respectively. In addition, the CMP distribution generalizes to the
geometric and Bernoulli distributions in the limiting cases
[\citet{shmueli2005useful}].

For the calculation of (\ref{eqcmppdf}), the $Z$-function needs to be
computed numerically due to the summation of an infinite series. For
certain combinations of $\lambda$ and $\nu$, many terms will be
needed in order to truncate the infinite summation with sufficient
accuracy, which leads to intensive computation. For these cases,
\citet{minka2003computing} derived an asymptotic approximation to
the $Z$-function, which is accurate when $\lambda> 10^{\nu}$.
\citet{wuCMP2013} discuss further improvements on computation by
taking advantage of parallel computing through Open Multiprocessing
(OpenMP) and Compute Unified Device Architecture (CUDA), that is,
graphics processing unit (GPU).

\section{Hierarchical Binomial mixture models}\label{secModeldev}
\subsection{Model development}

Let $\{\mathbf{s}_{i}\}_{i=1}^{G},\mathbf{s}_{i}\in D\subset\mbb{R}^{2}$
denote a set of sampling locations. We consider an experimental design
in which animals are surveyed at each sampling location $\mathbf{s}_{i}$
for a total of $J$ primary sampling occasions, in which there are
potentially $K$ nested secondary sampling periods. In principle, the
primary sampling occasions can be over any arbitrary time interval, for
example, in weeks or months. In addition, we assume a closed population
within each primary sampling occasion so that the species abundance at
each location varies across primary sampling occasions but not within.
Relative to the primary sampling occasion, the secondary sampling
period might be over a shorter time interval, for example, daily
surveys within the three-month long primary sampling occasions. To
allow for an unbalanced data structure, due to missing observations, we
assume $n_{ij}\leq K$ successful visits to site $\mathbf{s}_{i}$ during
the $j$th primary sampling period with the number of animals detected
recorded. Therefore, it follows that $0\leq n_{ij} \leq K$,
$i=1,2,\ldots,G$; $j=1,2,\dots,J$. We note that ``missing'' values are
not uncommon and can occur for many reasons. For example, some
scheduled visits might not be made due to illness of the observer, and
as a result no data will be recorded. In the current context, we assume
that any missing data are missing completely at random (MCAR)
[\citet{little2002statistical}].

For $i=1,2,\ldots,G$, $j=1,2,\ldots,J$, and $k=1,2,\ldots,n_{ij}$,
let $y_{ijk}$ be the number of animals observed at location $\mathbf
{s}_{i}$ during the $k$th secondary sampling within the $j$th primary
sampling occasion. The observed data can be denoted by $\mathbf{Y}=\{
\mathbf{y}_{ij}\dvtx i=1,2,\ldots,G; j=1,2,\ldots,J\}$, where $\mathbf
{y}_{ij}=(y_{ij1},y_{ij2},\ldots,y_{ijn_{ij}})'$ and $1 \leq n_{ij}
\leq K$. Note that $n_{ij}=0$ corresponds to the case that no
successful visits are made to site $i$ and, thus the vector $\mathbf
{y}_{ij}$ does not have any elements. Further, let $p_{ijk}$ be the
probability of detecting an animal during the $k$th ($k=1,2,\ldots
,n_{ij}$) secondary sampling within the $j$th primary sampling occasion
($j=1,2,\ldots,J$) at location $\mathbf{s}_{i}$ and denote $N_{ij}$ as
the unknown animal abundance at location $\mathbf{s}_{i}$ during the
$j$th primary sampling occasion. In other words, $N_{ij}$ represents
the total number of animals available for sampling during the $j$th
primary sampling occasion at location $\mathbf{s}_{i}$. Due to the closed
population assumption, $N_{ij}$ does not vary among secondary sampling
periods within each primary sampling occasion.

The nested design we consider is more general than many of the designs
previously investigated [e.g., \citet
{royle2004n,kery2005modeling,royle2005general,royle2006hierarchical,kery2008estimating,webster2008bayesian}],
all of which can be seen as a special case of ours by setting $K=1$. In
contrast, our study design is more similar to those found in
\citet{chandler2011inference} and \citet{dail2011models}.
Additionally, for the sake of flexibility, it is not necessary that
$n_{ij} \equiv K$ (for all $i=1,2,\ldots,G$ and $j=1,2,\ldots,J$).
Importantly, the replicated data collected in the secondary sampling
provides additional information that could alleviate potential issues
caused by missing values as well as improve the accuracy of parameter
estimation over the nonnested design. The primary objective of our
analysis is to estimate abundance and draw inference about
detectability. To achieve these goals, we propose a class of
hierarchical binomial mixture models, that includes the Bin-CMP model.

The class of binomial mixture models naturally fits into the
hierarchical framework [e.g., \citet
{royle2008hierarchical,cressiewikle2011}]. In this framework, we define
the \textit{observation model} as
%
\begin{equation}
y_{ijk}|N_{ij},p_{ijk}\sim\operatorname{Bin}(N_{ij},p_{ijk}),
\label{eqdatamodel}
\end{equation}
for $i=1,2,\ldots,G$; $j=1,2,\ldots,J$; $k=1,2,\ldots,n_{ij}$, where
the probability $p_{ijk}$ corresponds to the $k$th secondary sampling
within the $j$th primary sampling occasion at location $\mathbf{s}_i$.
For the design we consider, (\ref{eqdatamodel}) allows us to estimate
abundance parameters $N_{ij}$, which are both location- and
time-specific. Also, since the abundance $N_{ij}$ at each site $\mathbf
{s}_{i}$ varies over time, we are able to describe the temporal changes
in species abundance for all spatial locations, which is often vital in
the context of long-term ecological monitoring studies. Another benefit
of the design we consider is the potentially sharper estimates of the
detection probability. Using a single probabilistically coherent model,
we are able to provide spatial maps that illustrate the changes in
abundance over time as well as the spatial variation [e.g., see
Figures~2 and 3 in the supplementary article, \citet
{wuSupp2015}]. More importantly, (\ref{eqdatamodel}) also suggests how
over- and underdispersion can be explicitly accounted for in the
subsequent model development through the choice of an appropriate count
model for
abundance parameter, $N_{ij}$. Specifically, under the assumption of
independence between $N_{ij}$ and $p_{ijk}$, it follows that
\begin{eqnarray*}
\mathit{E}(y_{ijk})&=&E(p_{ijk})E(N_{ij}),
\label{eqyijkcondmu}
\\
\operatorname{Var}(y_{ijk})&=&E(p_{ijk})E(N_{ij})+E
\bigl(p_{ijk}^{2}\bigr)\bigl\{\operatorname{Var}(N_{ij})-E(N_{ij})
\bigr\}.
\end{eqnarray*}
Hence, the mean and variance relationship in the data can be addressed
through that of $N_{ij}$. For example, for data with over- and
underdispersion, we can choose a model for $N_{ij}$ such that
$\operatorname{Var}(N_{ij})>\mathit{E}(N_{ij})$ or $\operatorname
{Var}(N_{ij})<\mathit{E}(N_{ij})$, respectively. As such, our approach
addresses over- and underdispersed count data through the choice of an
appropriate model for abundance parameter, $N_{ij}$.

For $i=1,2,\ldots,G$ and $j=1,2,\ldots,J$, the \textit{process
model} we consider for the abundance, $N_{ij}$, is given by
%
\begin{equation}
N_{ij}|\lambda_{ij},\nu_{j} \sim f(
\lambda_{ij},\nu_{j}), \label{eqprocmodel}
\end{equation}
where $f(\cdot)$ is used to generically denote an appropriate count
distribution with intensity parameter $\lambda_{ij}$ and primary
sampling period-varying dispersion parameters $\nu_{j}$. There are
many possible choices for the distribution function $f(\cdot)$ in the
process model (\ref{eqprocmodel}), including the Pois, NB, and CMP,
among others. We focus on the case where $f(\cdot)$ is chosen to be
the CMP distribution, resulting in a flexible Bin-CMP mixture model
that allows for equi-, over-, and/or underdispersion. Alternatively, if
$f(\cdot)$ is chosen to be the NB distribution, the resulting Bin--NB
mixture model provides a suitable candidate for modeling overdispersed
data. Finally, it is important to note that, although we focus on the
CMP distribution, in our framework, $f(\cdot)$ can be chosen to be any
valid count distribution.

Specification of the \textit{parameter model} is usually
problem-specific and often depends on the research questions under
consideration. In long-term ecological monitoring studies, it is often
of interest to understand which factors might be important constituents
in the probability of detection, so that an efficient sampling protocol
can be designed. To achieve this goal, we relate the detection
probability, $p_{ijk}$, to the covariates $x_{ijk,1},\ldots,x_{ijk,P}$
through a logistic link function, that is,
%
\begin{equation}
\operatorname{logit}(p_{ijk})=\beta_{1}x_{ijk,1}+
\cdots+\beta_{P}x_{ijk,P}, \label{eqdetectmodel}
\end{equation}
where $\operatorname{logit}(r)=\log\{r/(1-r) \}$,
$i=1,2,\ldots,G$, $j=1,2,\ldots,J$, and
$k=1,2,\break \ldots,n_{ij}$. Note that (\ref{eqdetectmodel}) allows for an
intercept, by setting
$x_{ijk,1}\equiv1$ for all $i$, $j$, and $k$. By incorporating
covariates into the model,
the objective is to identify and draw statistical inference on
important factors governing the
probability of detection. Another interest in long-term ecological
studies is to gain deeper
understanding surrounding the intensity $\lambda_{ij}$, which
influences species abundance.
The second part of the \textit{parameter model} defines a model for
the intensity, $\lambda_{ij}$, as
%
\begin{eqnarray}\label{eqlocmumodel}
\log\lambda_{ij}=\mathbf{w}_{ij}'\bolds{\gamma}=w_{ij,1}\gamma
_{1}+\cdots+w_{ij,M}
\gamma_{M},
\nonumber\\[-8pt]\\[-8pt]
\eqntext{i=1,\ldots,G; j=1,\ldots,J.}
\end{eqnarray}
Here, $\mathbf{w}_{ij}=(w_{ij1},\ldots,w_{ij,M})'$ are a set of
covariates and $\bolds{\gamma}=(\gamma_{1},\ldots,\gamma_{M})'$
denotes the associated coefficients.

\subsection{Accounting for spatial dependence}
For spatially replicated count data, such as those typically
encountered in monitoring studies, it is sometimes necessary to
explicitly account for spatial dependence in the model for intensity.
Under this scenario, we can extend (\ref{eqlocmumodel}) to explicitly
incorporate spatial dependence by adding a spatial component in the
model for the intensity, that is,
%
\begin{equation}
\log\lambda_{ij}=\mathbf{w}_{ij}'\bolds{
\gamma}+\bolds{\phi}_{i}^{\prime}\bolds{\alpha}_{j},\qquad
i=1,\ldots,G; j=1,\ldots,J, \label{eqintensitymodel}
\end{equation}
or
\[
\log\bolds{\lambda}=\mathbf{w}'\bolds{\gamma}+\bigl(\bolds{
\Phi} \otimes\bolds{\alpha}'\bigr) \operatorname{vec}(
\mathbf{I}_{\tau\times\tau}),
\]
where $\bolds{\alpha}_{j}=(\alpha_{j1},\ldots,\alpha_{j\tau})'$;
$\bolds{\alpha}=(\bolds{\alpha}_{1}, \bolds{\alpha}_{2},\ldots,\bolds
{\alpha}_{J})$; $\bolds{\lambda}=(\lambda_{11},\ldots,\lambda
_{1J},\ldots,\lambda_{G1},\break  \ldots,\lambda_{GJ})'$; $\mathbf{w}=(\mathbf
{w}_{11},\ldots,\mathbf{w}_{1J},\ldots,\mathbf{w}_{G1},\ldots,\mathbf
{w}_{GJ})$; $\bolds{\Phi}$ denotes a $G \times\tau$ matrix of spatial
basis functions $\bolds{\Phi}=[\phi_{1}^{\prime}; \ldots; \phi
_{G}^{\prime}]$;
$\bolds{\phi}_{i}^{\prime}=(\phi_{i1},\ldots,\phi_{i\tau})$ is a row
vector denoting the $i$th row of $\bolds{\Phi}$; $\mathbf{I}_{\tau\times
\tau}$ is a $\tau\times\tau$ identity matrix; $\tau$ is the number
of basis functions and $\bolds{\alpha}\sim N(\mathbf{0},\bolds{\Sigma
}_\alpha)$. There are several advantages to incorporating spatial
effects when modeling the intensity function. Most importantly,
capturing spatial dependence in the intensity function among
neighboring locations will allow us to borrow strength from correlated
observations, potentially improving parameter estimation, statistical
inference, and prediction.

The choice of basis functions is typically problem specific, with
advantages arising from specific choices. Popular choices include
empirical orthogonal functions (EOFs), Fourier basis function, splines,
wavelets, bi-square and predictive process basis [e.g., see \citet
{royle2005efficient,cressie2008fixed,cressiewikle2011} and the
references therein]. In spatial statistical modeling, low-rank
representations are often considered [\citet{wikle2010}].
Following \citet{ruppert2003semiparametric}, we use the thin
plate spline basis functions, where
\[
\bolds{\Phi}=\mathop{\bigl[C(\mathbf{s}_{i}-\bolds{\kappa}_{l})
\bigr]}_{1\leq l \leq\tau}{}_{1 \leq i \leq I} \quad\mbox{and}\quad C(\mathbf{r})=\llVert\mathbf{r}
\rrVert^{2v-2}\log\llVert\mathbf{r}\rrVert,\qquad v>1,
\]
where $\bolds{\kappa}_{l}$ ($l=1,2,\ldots,\tau$) denote fixed knot
points in $\mbb{R}^{2}$ and $v$ is a smoothness parameter [see
\citet{holan2008semiparametric} for further discussion]. Here, we
choose $v=2$ [cf. \citet{ruppert2003semiparametric}, page 257]
and assume $\operatorname{cov}(\bolds{\alpha}_{j})=\sigma_{\alpha
_{j}}^{2}\bolds{\Omega}$, where
\[
\bolds{\Omega}=\mathop{\bigl[C(\bolds{\kappa}_{l}-\bolds{\kappa}_{l'})\bigr]}_{1 \leq l,l'
\leq\tau}.
\]
The selection of knot points can be facilitated through space-filling
designs, as implemented in the {\tt{fields}} package [\citet
{furrer2009fields}] in R [\citet{Rsoftware}]. The number of
knots $\tau$ can be chosen based on computational considerations
followed by sensitivity analysis. Alternatively, the number of knots
can be chosen according to $\tau=\max\{20, \min(G/4, 150)\}$
[\citet{ruppert2003semiparametric}, page 257]. Following\vspace*{2pt}
\citet{ruppert2003semiparametric}, we define $\bolds{\Phi
}^{*}=\bolds{\Phi}\bolds{\Omega}^{-1/2}$ and $\bolds{\alpha}^{*}=\bolds
{\Omega}^{1/2}\bolds{\alpha}$.
Then, for $i=1,2,\ldots,G$ and $j=1,2,\ldots,J$, we can rewrite (\ref
{eqintensitymodel}) as
%
\begin{equation}
\log\lambda_{ij}=\mathbf{w}_{ij}'\bolds{
\gamma}+\bolds{\phi}_{i}^{*\prime}\bolds{\alpha}_{j}^{*}=
\mathbf{g}_{ij}' \widetilde{\bolds{\gamma}}_{j},
\label{eqintensitymodelbasis}
\end{equation}
where $\bolds{\phi}_{i}^{*\prime}$ is the $i$th row of the matrix
$\bolds
{\Phi}^{*}$ and $\operatorname{cov}(\bolds{\alpha}_{j}^{*})=\sigma
_{\alpha_{j}}^{2}\mathbf{I}_{\tau\times\tau}$. Further, $\mathbf
{g}'_{ij}=(\mathbf{w}_{ij}' \bolds{\phi}_{i}^{*\prime})$ and $\widetilde
{\bolds{\gamma}}_{j}=(\gamma_{1},\ldots,\gamma_{M},\alpha
_{j1}^{*},\ldots,\alpha_{j \tau}^{*})'$.

\subsection{The likelihood}
To\vspace*{1.5pt} account for spatial dependence, we require that $\bolds{\alpha
}_{j}^{*}$, $j=1,2,\ldots,J$ in (\ref{eqintensitymodelbasis}) are in
the model with probability one. Since (\ref{eqlocmumodel}) and (\ref
{eqintensitymodelbasis}) are essentially of the same form, we will use
the former in the subsequent discussion. We now derive the likelihood
function for the model defined by (\ref{eqdatamodel}), (\ref
{eqprocmodel}), (\ref{eqdetectmodel}), and~(\ref{eqlocmumodel}). Let
$\mcal{M}=\{\mcal{M}_{\bolds{\beta}},\mcal{M}_{\bolds{\gamma}},\mcal
{M}_{\bolds{\nu}}\}$, and $\mcal{M}_{\bolds{\beta}},\mcal{M}_{\bolds
{\gamma}},\mcal{M}_{\bolds{\nu}}$ denote the model structures for the
set of covariates $\mathbf{x}$ and $\mathbf{w}$ and the dispersion
parameters $\bolds{\nu}=\{\nu_{1},\ldots,\nu_{J}\}$, respectively.
For example, in the case of $P=6, M=6,J=5$, $\mcal{M}_{\bolds{\beta
}}=\{x_{1},x_{3}\}$ indicates that only $x_{1}$ and $x_{3}$ are
included in the model for detection probability or, equivalently,
$\beta_{2}=\beta_{4}=\beta_{5}=\beta_{6}=0$; $\mcal{M}_{\bolds
{\gamma}}=\{w_{1},w_{2}\}$ indicates that only $w_{1}$ and $w_{2}$ are
included in the model for intensity; $\mcal{M}_{\bolds{\nu}}=\{
1,2,\ldots,J\}$ indicates that there is only one grouping for
dispersion parameters, meaning $\nu_{j}\equiv\nu$ for $j=1,2,\ldots,J$.
Under the assumption of conditional independence, the likelihood
function for the binomial mixture models we propose is given by
%
\begin{equation}
\mcal{L}(\mathbf{Y}|\mcal{M},\bolds{\beta},\bolds{\gamma},\bolds{\nu
})=\prod
_{i=1}^{G}\prod_{j=1}^{J}
\prod_{k=1}^{n_{ij}}{[y_{ijk}|N_{ij},
\bolds{\beta},\mcal{M}_{\bolds{\beta
}}]} {[N_{ij}|\mcal{M}_{\bolds{\gamma}},
\bolds{\gamma},\mcal{M}_{\bolds
{\nu}},\nu_{j}]},\hspace*{-15pt} \label{eqllikeyij}
\end{equation}
where, generically, $[\xi|\bolds{\theta}]$ denotes the conditional
distribution of $\xi$ given the parameters $\bolds{\theta}$.
Integrating out $N_{ij}$ in (\ref{eqllikeyij}) yields the marginal
distribution of observing $\mathbf{y}_{ij}$ as
%
\begin{eqnarray}\label{eqyijmarginal}
&& P(\mathbf{y}_{ij}|\mcal{M},\bolds{\beta},\bolds{\gamma},\nu_{j})\nonumber
\\
&&\qquad = \sum_{N_{ij}\geq y_{ij}^{\max}}^{\infty} \Biggl\{\prod_{k=1}^{n_{ij}}{\frac
{N_{ij}!}{y_{ijk}!(N_{ij}-y_{ijk})!}p_{ijk}^{y_{ijk}}(1-p_{ijk})^{N_{ij}-y_{ijk}}}
\Biggr\}
\\
&&\hspace*{66pt}{}\times f(N_{ij}|\mcal{M}_{\bolds{\gamma}},\bolds{\gamma},\mcal
{M}_{\bolds{\nu}},v_{j}), \nonumber
\end{eqnarray}
where $y_{ij}^{\max}=\max\{\mathbf{y}_{ij}\}$. Consequently, we can
derive the joint posterior distribution function $\pi(\mcal{M},\bolds
{\beta},\bolds{\gamma},\bolds{\nu}|\mathbf{Y})$ based on (\ref
{eqyijmarginal}) as
%
\begin{eqnarray}\label{eqintegratedllike}
&& \pi(\mcal{M},\bolds{\beta},\bolds{\gamma},\bolds{\nu}|\mathbf{Y})\nonumber
\\
&&\qquad \propto
\Biggl\{\prod_{i=1}^{G}\prod
_{j=1}^{J}{P(\mathbf{y}_{ij}|\mcal{M},
\bolds{\beta},\bolds{\gamma},\nu_{j})} \Biggr\}
\\
&&\hspace*{31pt}{} \times[\bolds{\beta}|\mcal{M}_{\bolds{\beta}}] [\bolds{\gamma}|\mcal{M}_{\bolds{\gamma}}]
[\bolds{\nu}| \mcal{M}_{\bolds{\nu}}] [\mcal{M}_{\bolds{\beta}}] [\mcal{M}_{\bolds
{\gamma}}] [\mcal{M}_{\bolds{\nu}}]. \nonumber
\end{eqnarray}
Here $[\bolds{\theta}]$ denotes the joint prior distribution function
of the parameters $\bolds{\theta}$.

Examination of (\ref{eqintegratedllike}) raises several computational
concerns. First, the calculation of $P(\mathbf{y}_{ij}|\mcal
{M},\bolds{\beta},\bolds{\gamma},\nu_{j})$ can be computationally
prohibitive, since a multiple integral is involved. This computational
issue becomes exacerbated when the domain of $N_{ij}$ covers a wide
range of values and/or if $G$ and $J$ are large. In addition to
calculating a multiple integral, in the case where $f(\cdot)$ denotes
the CMP distribution, evaluating (\ref{eqyijmarginal}) requires
computing the $Z$-function, which involves the summation of infinite
series. Specifically, for the Bin-CMP model, it is worth pointing out
that within each MCMC iteration, sampling elements in $\bolds{\gamma}$
or $\bolds{\nu}$ from their full conditionals requires both the
computation of the multiple integral and the approximation of the
$Z$-function. Therefore, implementation of our proposed model can be
computationally intensive in some cases. We resolve these computational
issues through the use of low level programming in C and parallel
computing with OpenMP.

Finally, we assume the following prior distributions for the model
parameters: $\bolds{\beta}\sim\operatorname{Gau}(\bolds{\mu}_{\beta
},\bolds{\Sigma}_{\beta})$; $\bolds{\gamma}\sim\operatorname
{Gau}(\bolds{\mu}_{\gamma},\bolds{\Sigma}_{\gamma})$. For the
dispersion parameters, we assume $\nu_{j}\sim\operatorname
{Unif}(a_{j},b_{j})$, $j=1,2,\ldots,J$, where $a_{j}$ and $b_{j}$ are
chosen appropriately to allow for different levels of dispersion in the
data (e.g., for overdispersed data, one may set $a_{j}\equiv0.02$ and
$b_{j}\equiv1.0$). In our case, we assign vague prior distributions
that are noninformative relative to scale of the data.

\section{Automated Bayesian model selection}\label{secModelsel}
For the binomial mixture models we propose, there are several
ecological objectives. First, there is a clear need to identify
important covariates among a set of candidate covariates in order to
gain an understanding of the factors affecting the detectability for a
given species of interest. In addition, the selection of influential
covariates is vital for studying which factors influence species
abundance. Last, the grouping of dispersion parameters will provide us
with further information about the level of dispersion associated with
the data across different years in the study. In such cases, grouping
is desired since some years may exhibit a similar level of dispersion
due to environmental changes or other exogenous factors. For example,
in our setting, specific neighborhoods may experience slow growth in
terms of the number of buildings established and/or certain climate
conditions may be more (or less) similar from year to year. Thus, it is
conceivable that some years may experience a similar dispersion
parameter. As such, we allow for data-driven grouping of the dispersion
parameters. To achieve these goals, we first discuss variable selection
and grouping in the context of the models we propose.

\subsection{Bayesian variable selection and grouping}
The literature on Bayesian variable selection is fairly extensive
[e.g., see \citet{o2009review,hooten2014guide} for a
comprehensive review]. Among the many available choices, the two most
commonly used techniques are stochastic search variable selection
[\citeauthor{george1993variable} (\citeyear{george1993variable,george1997approaches})] and reversible
jump MCMC \mbox{(RJMCMC)} [\citet{green1995reversible}]. For grouping,
however, \mbox{RJMCMC} is typically considered more appropriate and, thus, we
utilize it for both model selection and grouping. Although one could
consider model selection through various model selection criteria
[e.g., Deviance Information Criterion---\citet
{spiegelhalter2002bayesian}], this would be less advantageous when the
goal is both simultaneous variable selection and grouping.

For convenience of exposition, we explain our algorithm in the context
of the Bin-CMP model and note that the migration to other binomial
mixture models is analogous. The implementation of variable selection
for $\mathbf{x}$ and $\mathbf{w}$ involves two types of moves: BIRTH (B)
and DEATH (D) defined as follows:
\begin{enumerate}[B:]
\item[B:] propose to add a covariate ($x_m$ or $w_m$) to the current
model with probability $p_{m}^{b}$,

\item[D:] propose to remove a covariate ($x_m$ or $w_m$) from the
current model with probability $p_{m}^{d}$.
\end{enumerate}
As an example, we consider a D move for $\mathbf{x}$. In general, only a
subset of covariates are subject to variable selection, while others
are forced to remain in the model with probability one. For notational
simplification, let $\mathbf{A}_{x}$ denote the set of indices
corresponding to covariates $\mathbf{x}$ that are available for variable
selection. For example, if there are three covariates $x_{1}$, $x_{2}$,
and $x_{3}$ available and only $x_{1}$ and $x_{3}$ are subject to
variable selection (i.e., $x_{2}$ is in the model with the probability
1), then we have $\mathbf{A}_{x}=\{1,3\}$. Moreover, let $|\mathbf{A}_{x}|$
denote the cardinality of the set $\mathbf{A}_{x}$. For each covariate in
$\mathbf{A}_{x}$, we assume an equal probability of a B or D move, that is,
\[
p_{m}^{b}=p_{m}^{d}=1/2\qquad\mbox{for }m \in\mathbf{A}_{x}.
\]
Suppose at the current iteration $t$, the model structure is given by
$\mcal{M}^{t}=\{\mcal{M}_{\bolds{\beta}}^{t},\mcal{M}_{\bolds{\gamma
}}^{t},\mcal{M}_{\bolds{\nu}}^{t}\}$. The\vspace*{1pt} RJMCMC algorithm for
variable selection on $\mathbf{x}$ can be outlined as follows:
\begin{enumerate}[\textit{Step} 2:]
\item[\textit{Step} 1:] Start with the model structure $\mcal
{M}^{t}=\{\mcal{M}_{\bolds{\beta}}^{t},\mcal{M}_{\bolds{\gamma
}}^{t},\mcal{M}_{\bolds{\nu}}^{t}\}$, where $\mcal{M}_{\bolds{\beta
}}^{t}=\{x_{i_{1}},\ldots,x_{i_{m}}\}$ with $\bolds{\beta}^{t}=\{\beta
_{i_1},\ldots,\beta_{i_m}\}$.

\item[\textit{Step} 2:] Randomly draw an index from $\mathbf{A}_{x}$
with an equal probability $1/|\mathbf{A}_{x}|$. Assume $i_{s} \in\mathbf
{A}_{x}$ is chosen:
\begin{itemize}
\item[--] if $i_{s} \in\mcal{M}_{\bolds{\beta}}^{t}$, then propose a D
move and obtain $\mcal{M}_{\bolds{\beta}}^{\prime}=\mcal{M}_{\bolds
{\beta
}}^{t} \setminus\{x_{i_{s}}\}$ and $\mcal{M}^{\prime}=\{\mcal{M}_{\bolds
{\beta}}^{\prime},\mcal{M}_{\bolds{\gamma}}^{t},\mcal{M}_{\bolds{\nu
}}^{t}\}$ and $\bolds{\beta}^{\prime}=\{\beta_{i_1},\ldots,\beta
_{i_s}=0,\ldots,\beta_{i_m}\}$;
\item[--] otherwise propose a B move and obtain $\mcal{M}_{\bolds{\beta
}}^{\prime}=\mcal{M}_{\bolds{\beta}}^{t} \cup\{x_{i_{s}}\}$ and $\mcal
{M}^{\prime}=\{\mcal{M}_{\bolds{\beta}}^{\prime},\mcal{M}_{\bolds{\gamma
}}^{t},\mcal{M}_{\bolds{\nu}}^{t}\}$ and $\bolds{\beta}^{\prime}=\{\beta
_{i_1},\ldots,\beta_{i_m},\beta_{i_s}\}$.
\end{itemize}

\item[\textit{Step} 3:] Adjust the coefficient $\beta_{is}$
corresponding to the covariate $x_{i_{s}}$:
\begin{itemize}
\item[--] if a D move, set $\beta_{is}=0$;
\item[--] otherwise generate $\beta_{is}\sim q(\cdot)$.
\end{itemize}
\item[\textit{Step} 4:] Generate $u\sim\operatorname{Unif}(0,1)$:
\begin{itemize}
\item[--] if $u< \min\{1,\mathrm{BF}(\mcal{M}_{\bolds{\beta}}^{\prime},\mcal
{M}_{\bolds{\beta}}^{t})\times R\}$, then set $\mcal{M}_{\bolds{\beta
}}^{t+1}=\mcal{M}_{\bolds{\beta}}^{\prime}$ and\break $\mcal{M}^{t+1}=\mcal
{M}^{\prime}$;
\item[--] otherwise $\mcal{M}_{\bolds{\beta}}^{t+1}=\mcal{M}_{\bolds{\beta
}}^{t}$ and $\mcal{M}^{t+1}=\mcal{M}^{t}$.
\end{itemize}
\item[\textit{Step} 5:] Repeat.
\end{enumerate}
In terms of the proposal distribution $q(\cdot)$, we used a
$\operatorname{Gau}(0,\zeta)$ distribution with $\zeta$ being a
user-defined tuning parameter. Moreover,
\[
R= \cases{
\displaystyle\frac{p_{i_{s}}^{b}}{p_{i_{s}}^{d}}\times q(\beta_{is}),
&\quad if D move,
\vspace*{5pt}\cr
\displaystyle\frac{p_{i_{s}}^{d}}{p_{i_{s}}^{b}}\times\frac{1}{q(\beta_{is})},
&\quad if B move,}
\]
and
\[
\operatorname{BF}\bigl(\mcal{M}_{\bolds{\beta}}^{\prime},
\mcal{M}_{\bolds{\beta
}}^{t}\bigr)=\frac{P(\mcal{M}_{\bolds{\beta}}',\bolds{\beta}'|\mathbf
{Y},\mcal{M}_{\bolds{\gamma}}^{t},\bolds{\gamma},\mcal{M}_{\bolds{\nu
}}^{t},\bolds{\nu})}{P(\mcal{M}_{\bolds{\beta}}^{t},\bolds{\beta
}^{t}|\mathbf{Y},\mcal{M}_{\bolds{\gamma}}^{t},\bolds{\gamma},\mcal
{M}_{\bolds{\nu}}^{t},\bolds{\nu})}.
\]

We now discuss the grouping algorithm for the dispersion parameters
$\bolds{\nu}$. Assume there are $n_{t}$ different arrangements
$T_{1},T_{2},\ldots,T_{n_{t}}$ for $\bolds{\nu}$ at the $t$th
iteration of the MCMC, that is, $\mcal{M}_{\bolds{\nu}}^{t}=\{T_{1},
T_{2}, \ldots, T_{m}, \ldots,T_{n_{t}}\}$. For each grouping $T_{m}$,
$m=1,2,\ldots,n_{t}$, the corresponding elements are subscripts for
the dispersion parameter group membership. For example, if $n_{t}=1$,
we have $T_{1}=\{1,2,\ldots,J\}$, that is, $\nu_{j} \equiv\nu$, for
$j=1,2,\ldots,J$. Similar to the variable selection previously
described, we allow for two types of moves as follows:
\begin{enumerate}[C:]
\item[C:] propose to combine two different arrangements into one
arrangement with $p_{c}$,
\item[S:] propose to split the arrangement into two arrangements with
probability $p_{s}$.
\end{enumerate}
Without loss of generality, assume an equal probability of proposing a
C or S move, that is, $p_{c}=p_{s}=1/2$. As an illustration, we
describe only the S move. Suppose there are $n_{t}^{s}$ out of $n_{t}$
arrangements in $\mcal{M}_{\bolds{\nu}}^{t}$ that have more than one
single element. We randomly choose each of these $n_{t}^{s}$
arrangements with an equal probability. Assume that group $T_{m}$ is
chosen, where $m \in\{1,\ldots,n_{t}^{s}\}$ and $|T_{m}|>1$. Assuming
we split $T_{m}$ into two nonempty sets $T_{m_{1}}$ and $T_{m_{2}}$, we
denote the resulting model structure as $\mcal{M}_{\bolds{\nu}}'=\{
T_{1}, T_{2}, \ldots, T_{m_{1}}, T_{m_{2}}, \ldots,T_{n_{t}}\}$. The
RJMCMC algorithm for grouping of $\bolds{\nu}$ can be outlined as follows:
\begin{enumerate}[\textit{Step} 3:]
\item[\textit{Step} 1:] Calculate the probability $P(\mcal{M}_{\bolds
{\nu}}'|\mcal{M}_{\bolds{\nu}})$ and $P(\mcal{M}_{\bolds{\nu}}|\mcal
{M}_{\bolds{\nu}}')$ as
\begin{eqnarray*}
P\bigl(\mcal{M}_{\bolds{\nu}}'|\mcal{M}_{\bolds{\nu}}\bigr)&=&
\frac{1}{2}\frac
{1}{n_{t}^{s}}\frac{1}{2^{(|T_{m}|-1)}-1},
\\
P\bigl(\mcal{M}_{\bolds{\nu}}|\mcal{M}_{\bolds{\nu}}'\bigr)&=&
\frac{1}{2}\frac
{1}{{n_{t}^{s}+1 \choose2}}
\end{eqnarray*}
[\citet{king2002bayesian}].

\item[\textit{Step} 2:] Let $\nu_{m}$ denote the value common to
all dispersion parameters in $T_{m}$ and $\nu_{m_{1}}$ and $\nu
_{m_{2}}$ be the values of dispersion parameters in $T_{m_{1}}$ and $T_{m_{2}}$,
respectively. Define a bijective mapping between $\nu_{m}$ and $\nu
_{m_{1}},\nu_{m_{2}}$ as
\[
\nu_{m_{1}}=\nu_{m}+\varepsilon\quad\mbox{and}\quad
\nu_{m_{2}}=\nu_{m}-\varepsilon,
\]
where $\varepsilon\sim h(\cdot)$.
\item[\textit{Step} 3:] Generate $\xi\sim\operatorname{Unif}(0,1)$:
\begin{itemize}
\item[--] if $\xi< \min\{1,\mathrm{BF}(\mcal{M}_{\bolds{\nu}}^{\prime},\mcal
{M}_{\bolds{\nu}}^{t})\times R_{s}\}$, then set $\mcal{M}_{\bolds{\nu
}}^{t+1}=\mcal{M}_{\bolds{\nu}}^{\prime}$ and\break $\mcal{M}^{t+1}=\mcal
{M}^{\prime}$;

\item[--] otherwise $\mcal{M}_{\bolds{\nu}}^{t+1}=\mcal{M}_{\bolds{\nu
}}^{t}$ and $\mcal{M}^{t+1}=\mcal{M}^{t}$.
\end{itemize}
\end{enumerate}
In terms of the proposal distribution $h(\cdot)$, we used $h(\eta
)=\operatorname{Unif}(-\eta,\eta)$ where $\eta$ is chosen through
pilot tuning. Moreover,
\begin{eqnarray*}
\mathrm{BF}\bigl(\mcal{M}_{\bolds{\nu}}^{\prime},\mcal{M}_{\bolds{\nu
}}^{t}
\bigr)&=&\frac{P(\mcal{M}_{\bolds{\nu}}',\nu_{m_1},\nu_{m_2}|\mathbf
{Y},\mcal{M}_{\bolds{\gamma}}^{t},\bolds{\gamma},\mcal{M}_{\bolds
{\beta}}^{t},\bolds{\beta}^{t})}{P(\mcal{M}_{\bolds{\nu}}^{t},\nu
_{m}|\mathbf{Y},\mcal{M}_{\bolds{\gamma}}^{t},\bolds{\gamma},\mcal
{M}_{\bolds{\beta}}^{t},\bolds{\beta}^{t})},
\\
R_{s}&=& \frac{P(\mcal{M}_{\bolds{\nu}}|\mcal{M}_{\bolds{\nu
}}')}{P(\mcal{M}_{\bolds{\nu}}'|\mcal{M}_{\bolds{\nu}})} \times\frac
{1}{h(\varepsilon)}\times\biggl
\llvert{\frac{\partial{(\nu_{m_{1}},\nu
_{m_{2}})}}{\partial{(\nu_{m},\varepsilon)}}} \biggr\rrvert.
\end{eqnarray*}

\section{Simulated examples}\label{secSim}
To evaluate the performance of the binomial mixture models we propose,
we considered two simulated examples using the Bin-CMP model, the
difference of which only resides in whether or not a spatial component
is included in the intensity model. For both simulations, we choose
$G=131$, \mbox{$J=5$}, and $K=3$ to be the same as the American Robin data
presented in Section~\ref{secApp}. For both examples, we simulate data
as $y_{ijk}|N_{ij},p_{ijk} \sim
\operatorname{Bin}(N_{ij},p_{ijk})$. For the probability of detection,
we consider
\[
\operatorname{logit}(p_{ijk})=\beta_{1}x_{ijk,1}+
\beta_{2}x_{ijk,2}+\cdots+\beta_{P}x_{ijk,P},
\]
where the values for the covariates $\mathbf{x}$ are set to be the same
as in the American Robin data for $i=1,2,\ldots, G$, $j=1,2,\ldots,J$,
$k=1,2,\ldots,K$, $l=1,2,\ldots,$ $P=4$. In addition, we set $\bolds
{\beta}=(-2.31,-0.4,0.0,-0.4)'$ with $ \{x_{1},x_{2},x_{4}
\}$ being important covariates. For the true abundance parameters
$N_{ij}$, we simulated from $N_{ij}\sim\operatorname{CMP}(\lambda_{ij},\nu
_{j})$, with $\nu_{1}=\nu_{3}=\nu_{5}=0.15$, $\nu_{2}=\nu
_{4}=0.06$ and $\bolds{\gamma}_{0}=(0.31,0.13,0.44,0.16,0.35)'$, as
estimated from the American Robin data presented in Section~\ref
{secApp}. For $i=1,2,\ldots,G$ and $j=1,2,\ldots,J$, the intensity
$\lambda_{ij}$ is simulated according to
\begin{eqnarray*}
\mbox{\textbf{S1}:}\quad \log\lambda_{ij}&=&\mathbf{w}_{i}^{\prime}
\bolds{\gamma}+\gamma_{0j},
\\
\mbox{\textbf{S2}:}\quad \log\lambda_{ij}&=&\mathbf{w}_{i}^{\prime}
\bolds{\gamma}+\bolds{\phi}_{i}^{*\prime}\bolds{\alpha}+
\gamma_{0j},
\end{eqnarray*}
where $\bolds{\phi}_{i}^{*\prime}$ for $i=1,2,\ldots,G$ and $\bolds
{\gamma
}_{0}=(\gamma_{01},\ldots,\gamma_{05})'$ are determined according to
the American Robin data with $\tau=10$. In each of the two models,
$\mathbf{w}_{i}$ are set to be the same as in the American Robin data
presented in Section~\ref{secApp}. Further, we set $M=11$ and $\bolds
{\gamma}=(0.0,0.0,0.0,0.0,0.0,0.06,0.0,0.0,0.0,0.03,0.0)'$, that is,
with $ \{w_{6},w_{10} \}$ being important covariates.
Particularly, for \textbf{S2}, the coefficients of spatial components,
$\bolds{\alpha}$, are randomly sampled from $\operatorname{Unif}(0,1)$
to avoid $y_{ijk}$ being too large. For the two simulations, we apply
RJMCMC to perform variable selection and grouping. Similar to the
analysis presented in Section~\ref{secApp}, we require $\bolds{\alpha
}$ to be included in the model with probability one for \textbf{S2}
and set $a_{j}\equiv0.02$ and $b_{j} \equiv2.0$ to allow for both
over- and underdispersion. In addition, we set $\bolds{\mu}_{\beta
}=\bolds{\mu}_{\gamma}\equiv\mathbf{0}$, $\bolds{\Sigma}_{\beta
}=10^{2}\mathbf{I}_{P}$, and $\bolds{\Sigma}_{\gamma}=10^{2}\mathbf{I}_{M}$.

%
\begin{table}
\tabcolsep=0pt
\tablewidth=250pt
\caption{Posterior marginal probabilities of the most probable model
for $\mathbf{x}$, $\mathbf{w}$, and $\bolds{\nu}$ in the Bin-CMP mixture
models \textup{\textbf{S1}} and \textup{\textbf{S2}} simulated examples
(Section~\protect\ref{secSim}) using RJMCMC. Note that \textup{\textbf{S1}}
contains only the covariates in the intensity model, whereas \textup{\textbf{S2}} contains both covariates and spatial components in the intensity
model and that the posterior probability for $\mathbf{x}$ under both
\textup{\textbf{S1}} and \textup{\textbf{S2}} are slightly less than 1.00 and become
1.00 as a result of rounding}\vspace*{8pt}\label{tabBinMixSimRJMCMC}
(a) Variable selection and grouping for \textbf{S1}
\begin{tabular*}{\tablewidth}{@{\extracolsep{\fill}}@{}llcc@{}}
\hline
\textbf{Para-}& & & \textbf{Posterior}\\
\textbf{meter} & \textbf{Model} & \textbf{Frequency} & \textbf{probability}\\
\hline
$\mathbf{x}$ & $ \{x_{1},x_{2},x_{4} \}$ & 59,838 & 1.00 \\[3pt]
$\mathbf{w}$ & $ \{w_{6},w_{10} \}$ & 53,951 & 0.90 \\
& $ \{w_{2},w_{6},w_{10} \}$ & \phantom{0,}4386 & 0.07 \\[3pt]
$\bolds{\nu}$ & $T_{1}=\{2,4\},T_{2}=\{1,3,5\}$ & 43,507 & 0.73 \\
& $T_{1}=\{1,3\},T_{2}=\{2,4\}, T_{3}=\{5\}$ & \phantom{0,}7801 & 0.13 \\
& $T_{1}=\{1\},T_{2}=\{2,4\}, T_{3}=\{3,5\}$ & \phantom{0,}3918 & 0.07 \\
\hline
\end{tabular*}\vspace*{18pt}

(b) Variable selection and grouping for \textbf{S2}

\begin{tabular*}{\tablewidth}{@{\extracolsep{\fill}}@{}llcc@{}}
\hline
\textbf{Para-}& & & \textbf{Posterior}\\
\textbf{meter} & \textbf{Model} & \textbf{Frequency} & \textbf{probability}\\
\hline
$\mathbf{x}$ & $ \{x_{1},x_{2},x_{4} \}$ & 59,741 & 1.00 \\[3pt]
$\mathbf{w}$ & $ \{w_{6},w_{10} \}$ & 56,139 & 0.94 \\[3pt]
$\bolds{\nu}$ & $T_{1}=\{2,4\},T_{2}=\{1,3,5\}$ & 37,071 & 0.76 \\
& $T_{1}=\{3\}, T_{2}=\{2,4\},T_{2}=\{1,5\}$ & \phantom{0,}7573 & 0.13 \\
\hline
\end{tabular*}
\end{table}

Table~\ref{tabBinMixSimRJMCMC} provides the posterior marginal
probabilities for the most probable model for $\mathbf{x}$, $\mathbf{w}$,
and $\bolds{\nu}$ in the Bin-CMP models \textbf{S1} and \textbf{S2}.
For model \textbf{S1}, the most frequent detection probability model
was given by $ \{x_{1},x_{2},x_{4} \}$ and appeared with a
frequency of 99.73\%. The most frequent intensity model was defined by
$ \{w_{6},w_{10} \}$ and had a frequency of 89.92\%. In
addition, the most frequent grouping for dispersion parameters is
$\mcal{M}_{\bolds{\nu}}=\{\{2,4\},\{1,3,5\}\}$, which appeared with a
frequency of 72.51\%. In all cases, the RJMCMC correctly identified the
set of important covariates as well as grouping for dispersion
parameters with the posterior marginal probability greater than or
equal to 72.51\%. In terms of parameter estimation, in most cases the
95\% credible intervals (CIs), averaged over the different models,
contain the true values---providing further indication that the
correct model is selected with high probability. For model \textbf
{S2}, the most frequent set of covariates for the detection probability
model was given by $ \{x_{1},x_{2},x_{4} \}$ and appeared
with a frequency of 99.57\%. The most frequent set of covariates $
\{w_{6},w_{10} \}$ for the intensity model had a frequency of
93.57\%. In addition, the most frequent grouping for the dispersion
parameters is $\mcal{M}_{\bolds{\nu}}=\{\{2,4\},\{1,3,5\}\}$, which
appeared with a frequency of 76.00\%.

In summary, the two simulations suggest that we are able to correctly
identify important covariates and grouping for dispersion parameters
with high posterior probability. Finally, for the estimation of
abundance in the two simulations, our approach performs satisfactorily,
as measured by coverage of the 95\% CIs. In the presence of spatial
components, however, we note that the model averaged estimates of
dispersion parameters can be adversely affected by missing data.

\section{Application: The Baltimore Ecosystem Study}\label{secApp}

In the urban ecosystems literature, bird communities are often used as
surrogates for studying urban biodiversity or species responses to
urbanization [\citet{shochat2010birds,aronson2014global}]. Within
urban areas the bird community is shaped by local-scale features such
as habitat features that vary among neighborhoods, landscape pattern,
and socioeconomic characteristics of residents that may influence land
management decisions [\citet{pickett2012bio}]. The American
Community Survey (ACS) is an ongoing survey that is able to provide
timely economic, social, and demographic information on small
geographies such as census tracts. Thus, to examine the effects of
certain demographic characteristics on abundance, we consider several
ACS variables. Additionally, environmental features of different
neighborhoods can be described by many factors, such as vegetation
diversity and are, therefore, also considered in our analysis.

Substantial research has been undertaken to investigate how
socioeconomic status and environmental variables influence the
abundance and diversity of various avian species [see \citet
{loss2009relationships,smallbone2011anuran,denison2010effects} and the
references therein]. Using socioeconomic variables from the decennial
census in 2000 associated with each census tract block groups as
covariates, \citet{denison2010effects} considered a simple NB
regression with no spatial components under the frequentist paradigm to
estimate the relative abundance for European starling in the City of
Baltimore, Maryland using a portion of data collected from 2005 to
2007. In contrast, we consider American Robin data from the BES
collected from 2005 to 2009 and apply various Bin-CMP models in order
to select important covariates for estimating the detection probability
and abundance of the American Robin, as well as to identify the
grouping of dispersion parameters. Due to missing values, the data we
consider has an unbalanced structure. In particular, the percentage of
secondary sampling occasions with at least one missing observation for
each of five primary sampling occasions is 6.87\%, 6.87\%, 3.05\%,
77.1\%, and 50.38\%, respectively. Moreover, the overall percentage of
missing observations in the American Robin data set is 9.62\%.

For the American Robin data, a total of 131 bird survey points were
visited during three secondary daily surveys within each of the five
primary sampling occasions from 2005 to 2009. With three covariates
available, we considered a full model for the detection probability as
%
\begin{equation}
\operatorname{logit}(p_{ijk})=\beta_{1}+\beta_{2}
\texttt{time}_{ijk}+\beta_{3}\texttt{airtemp}_{ijk}+
\beta_{4}\texttt{cloudcover}_{ijk},\hspace*{-20pt} \label{eqndetectprobeust}
\end{equation}
for $i=1,\ldots, 131$, $j=1,\ldots, 5$, and $k=1,\ldots, n_{ij} \leq
K=3$. Regarding the covariates in (\ref{eqndetectprobeust}), \texttt
{time}, \texttt{airtemp}, and \texttt{cloudcover} correspond to the
start time, air temperature, and cloud cover (i.e., the fraction of the
sky obscured by clouds) recorded on each visit to the bird survey
points, respectively.

In terms of full models for the intensity, we considered the following
three models:
\begin{eqnarray*}
\mbox{\textbf{M1}:}\quad \log\lambda_{ij}&=&\mathbf{w}_{i}^{\prime}
\bolds{\gamma}+\widetilde{\bolds{\phi}}{}_{i}^{*\prime}\bolds{
\alpha}+\gamma_{0j},
\\
\mbox{\textbf{M2}:}\quad \log\lambda_{ij}&=&\mathbf{w}_{i}^{\prime}
\bolds{\gamma}+\gamma_{0j},
\\
\mbox{\textbf{M3}:}\quad \log\lambda_{ij}&=&\bolds{\phi}_{i}^{*\prime}
\bolds{\alpha}+\gamma_{0j}, \label{eqnintensityM2}
\end{eqnarray*}
where, for $j=1,\ldots,J$, $\gamma_{0j}$ is a year-specific intercept
and $\bolds{\phi}_{i}^{*\prime}$ is the $i$th row of the matrix $\bolds
{\Phi
}^{*}$ as discussed in Section~\ref{secModeldev}. Moreover, the
covariates in the intensity model are given by $\mathbf
{w}_{i}^{\prime}=(\texttt{uftree}_{i}$, $\texttt
{ufbldg}_{i}$, $\texttt{ufmgrass}_{i}$, $\texttt{bld200m}_{i}$, $\texttt{for200m}_{i}$,
$\texttt{veg200m}_{i}$, $\texttt{African}_{i}$, $\texttt{bachelor}_{i}$, $\texttt{fmkds}_{i}$, $\texttt{pubassit}_{i}$,
$\texttt{houseyr}_{i})$. These
covariates are specific to each survey location and do not vary with
primary sampling occasions. Among these environmental variables,
\texttt{uftree}, \texttt{ufbldg}, and \texttt{ufmgrass} are the
UFORE plots variables that indicate tree cover, ground cover by
buildings and maintained grass, respectively. Further, \texttt
{bld200m}, \texttt{for200m}, and \texttt{veg200m} are variables that
measure tree cover, other vegetation cover, and cover by buildings in
the 200 meter radius plot, respectively [see Figure~1 in the
supplemental article, \citet{wuSupp2015}]. For the ACS variables
specific to each census tract block group, \texttt{African} is the
percentage of African American residents; \texttt{bachelor} is the
percentage of population with Bachelor's degree or higher; \texttt
{fmkds} is the percentage of housing units occupied by female
householder and children under 18 years; \texttt{pubassit} is the
percentage of households on government public income assistance;
\texttt{hourseyr} is the median year that a housing unit was built. We
used the five-year period estimates from 2005 to 2009 for these ACS
variables, which can be obtained at the U.S. Census Bureau website
(\surl{http://www.census.gov/acs/www/}). Our specific choice of ACS
variables was facilitated by a social areas analysis approach
[\citet
{denison2010effects,maloney1974social,muller2013patterns}]. Note that
we standardize the covariates in (\ref{eqndetectprobeust}) and in the
intensity model for numerical stability. Further, based on exploratory
analysis involving various collinearity diagnostics (e.g., condition
number, etc.) of the site covariates (not shown) and subject matter
knowledge, we expect any effects of collinearity between the site
covariates to have a minimal affect on the variable selection
algorithm. Finally, for model \textbf{M1}, we orthogonalize the matrix
of spatial basis function with respect to covariates, to alleviate
potential confounding with the covariate effects [\citet
{hodges2010adding}]. As a result, $\widetilde{\bolds{\phi}}{}_{i}^{*\prime}$
is the $i$th row of the matrix of $\widetilde{\bolds{\Phi}}{}^{*}$ after
the orthogonalization.

It is worth pointing out that the choice of models above depends on the
goal of the ecological study. For example, \textbf{M3} can be used if
no covariates are available for modeling the intensity. For other cases
where covariates are available, but there is no spatial dependence (or
the spatial dependence is negligible after accounting for covariates),
model \textbf{M2} can be utilized. Given both covariates are available
and spatial dependence is present, \textbf{M1} represents a potential model.

When implementing the RJMCMC algorithm, we require the ``intercept''
term~$\beta_{1}$ in (\ref{eqndetectprobeust}) and $\bolds{\gamma
}_{0}$, in the model for intensity, to be included with probability
one. In addition, in the presence of spatial components, we require
$\bolds{\alpha}$ to be in the model for the intensity with probability
one. For the choice of knot points, when using low-rank thin plate
basis functions, we considered a sensitivity analysis to choose the
number of knots and a space-filling design for placement. Specifically,
for three different choices of the number of knot points, $\tau=10$,
15, and 32 in \textbf{M1}, similar results are obtained in terms of
abundance estimation, although parameter estimation becomes more
difficult as $\tau$ gets large. Equally important, the results of a
sensitivity analysis indicate that the variable selection and grouping
for the dispersion parameters seem robust to a different number of knot
points. Hence, we choose $\tau=10$ for both \textbf{M1} and \textbf
{M3}. We used a Metropolis--Hastings within Gibbs sampler consisting of
a total of 120,000 MCMC iterations, with the first 60,000 discarded as
burn-in. Our inference is based on every third sample after burn-in,
which results in a total of 20,000 samples used.

In terms of posterior marginal probability, the model having \texttt
{time} and \texttt{cloudcover} has the highest probability of being
selected in the model for detectability. Similarly, for the intensity
model, \texttt{ufbldg}, \texttt{veg200m}, and \texttt{pubassit} are
selected with higher probability relative to other covariates. However,
the grouping of dispersion parameters varies across models depending on
whether spatial components are included. This is not unexpected, as
there is a trade-off between the dispersion parameter and inclusion of
spatial components. The three models we considered all produce similar
results in terms of the selection of important covariates and abundance
estimates (results not shown). However, since the goal of our analysis
is to identify and draw inference on important covariates relating to
detectability and abundance, we present results from the more
parsimonious model~\textbf{M2}. From Table~\ref{tabBinCMPamroM2}, it
can be seen that \texttt{time} and \texttt{cloudcover} are identified
as important predictors for detectability of American Robin. For the
covariates in the intensity model, \texttt{ufbldg}, \texttt{veg200m},
and \texttt{pubassit} are selected as the important factors in all
cases. For the dispersion parameters, the results suggest the most
probable model has the grouping $T_{1}=\{2,4\},T_{2}=\{1,3,5\}$ (with
posterior probability 0.6496), indicating that the data in 2005, 2007,
and 2009 exhibit a similar amount of dispersion, whereas the data for
2006 and 2008 show similar amounts of dispersion.

%
\begin{table}
\tabcolsep=0pt
\tablewidth=250pt
\caption{Posterior probabilities of the most
probable model for \textup{\textbf{M2}} and the posterior summary statistics in
the Bin-CMP model assuming the posterior mode model for \textup{\textbf{M2}}.
Note that \textup{\textbf{M2}} only contains covariates in the intensity model,
and $\widehat{R}$ refers to the Gelman--Rubin diagnostics}\vspace*{8pt}\label{tabBinCMPamroM2}
(a) Variable selection and grouping
\begin{tabular*}{\tablewidth}{@{\extracolsep{\fill}}@{}llcc@{}}
\hline
& & & \textbf{Posterior} \\
\textbf{Variable} & \textbf{Model} & \textbf{Frequency} & \textbf{probability} \\
\hline
$\mathbf{x}$ &\{cloudcover\} & 31,992 & 0.53 \\
&\{time, cloudcover\} & 27,587 & 0.46 \\[3pt]
$\mathbf{w}$ &\{veg200m, pubassit\} & 51,343 & 0.86 \\
&\{ufbldg, veg200m, pubassit\} & \phantom{0,}7234 & 0.12 \\[3pt]
$\bolds{\nu}$ & $T_{1}=\{2,4\},T_{2}=\{1,3,5\}$ & 38,973 & 0.65 \\
& $T_{1}=\{2\},T_{2}=\{1,3,4,5\}$ & \phantom{0,}7445 & 0.12 \\
&$T_{1}=\{2\},T_{2}=\{4\},T_{2}=\{1,3,5\}$& \phantom{0,}3745 & 0.06 \\
\hline
\end{tabular*}\vspace*{18pt}

(b) Parameter estimation

\begin{tabular*}{\tablewidth}{@{\extracolsep{\fill}}@{}ld{2.2}d{2.2}d{2.2}d{2.2}c@{}}
\hline
\textbf{Parameter} & \multicolumn{1}{c}{$\bolds{\mu_{\mathrm{post}}}$} & \multicolumn{1}{c}{$\bolds{\sigma_{\mathrm{post}}}$}
& \multicolumn{1}{c}{$\bolds{Q_{0.025}}$} & \multicolumn{1}{c}{$\bolds{Q_{0.975}}$} & $\bolds{\widehat{R}}$ \\
\hline
intercept & -2.31 & 0.07 & -2.45 & -2.17 & 1.00\\
time & -0.10 & 0.03 & -0.15 & -0.04 & 1.00\\
cloudcover & -0.04 & 0.03 & -0.09 & 0.01 & 1.00\\
ufbldg & -0.02 & 0.01 & -0.03 & -0.01 & 1.00 \\
veg200m & 0.06 & 0.01 & 0.05 & 0.09 & 1.00\\
pubassit & 0.02 & 0.01 & 0.01 & 0.04 & 1.00\\
$\gamma_{01}$ & 0.35 & 0.07 & 0.23 & 0.51 & 1.01 \\
$\gamma_{02}$ & 0.16 & 0.05 & 0.07 & 0.26 & 1.01 \\
$\gamma_{03}$ & 0.48 & 0.08 & 0.33 & 0.67 & 1.01 \\
$\gamma_{04}$ & 0.14 & 0.05 & 0.05 & 0.24 & 1.01\\
$\gamma_{05}$ & 0.38 & 0.07 & 0.25 & 0.55 & 1.01 \\
$\nu_{24}$ & 0.08 & 0.02 & 0.05 & 0.11 & 1.01 \\
$\nu_{135}$ & 0.17 & 0.03 & 0.12& 0.23 & 1.01 \\
\hline
\end{tabular*}
\end{table}

Last, we consider the posterior mode model (i.e., the model with the
highest posterior probability) for the Bin-CMP mixture model \textbf
{M2} in order to draw inference about how the different covariates
affect high detectability and abundance of the American Robin within
the study domain. We conclude that an important covariate is a
positively (or negatively) significant factor if the lower (or upper)
end of 95\% CIs is greater (or smaller) than 0, respectively. For the
posterior mode model, we include only the intercept, \texttt{time},
and \texttt{cloudcover} in (\ref{eqndetectprobeust}), whereas for the
covariates in the intensity model, only \texttt{ufbldg}, \texttt
{veg200m}, and \texttt{pubassit} are included. For the dispersion
parameters, we consider the case where $\nu_{2}=\nu_{4}=\nu_{24}$
and $\nu_{1}=\nu_{3}=\nu_{5}=\nu_{135}$. Table~\ref
{tabBinCMPamroM2} presents the posterior summary statistics and
Gelman--Rubin diagnostics [\citet{brooks1998general}] for model
parameters.
It is shown that in all cases $\widehat{R}$ is close to 1,
indicating convergence has been reached. Moreover, \texttt{time} is
negatively correlated with the detectability of the American Robin,
that is, the earlier the survey is conducted, the more likely it is
that we can detect American Robin. In terms of the intensity, \texttt
{ufbldg} is negatively related to the abundance of American Robin,
whereas \texttt{veg200m} and \texttt{pubassit} are positively
related. As a result, for bird survey points nearby more buildings, the
abundance of American Robin is lower; while for survey points with a
higher percentage of vegetation and residents of lower socio-economic
status, the abundance of American Robin is higher.
%
%
%
\begin{figure}
\includegraphics{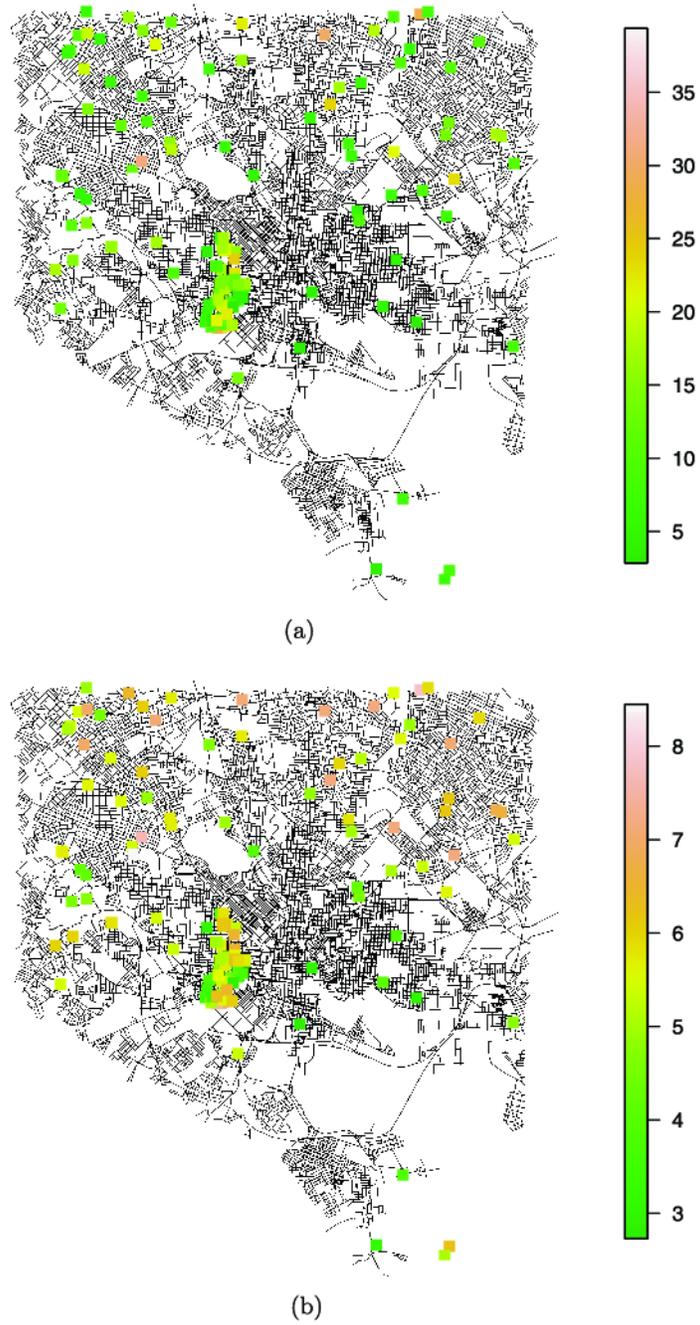}
\caption{Plots of posterior mean and standard deviation of abundance
estimates for 2009 in the Bin-CMP model assuming the posterior mode
model for \textup{\textbf{M2}}. Note that \textup{\textbf{M2}} only contains covariates
in the intensity model. \textup{(a)}~Posterior mean, \textup{(b)}~posterior sd.}\label{figBinCMPamroM2N2009}\label
{figY2009mu}\label{figY2009sd}
\end{figure}
As an example,
Figure~\ref{figBinCMPamroM2N2009} provides a spatial map for the
posterior mean and standard deviation of the abundance estimate (from
\textbf{M2}) for 2009, whereas Figures~2 and 3 of the supplemental
article [\citet{wuSupp2015}] illustrate how the abundance
estimates and their standard errors change over the duration of the
period studied (2005--2009). Last, our results suggest that the
American Robin are overdispersed within the study domain over all of
the years considered.

\section{Discussion}\label{secDiscu}
Motivated by the American Robin data from the BES, we developed a class
of Bayesian hierarchical binomial mixture models that allow for
automated variable selection and grouping in the presence of unbalanced
nested design. In addition, we demonstrate that over- and
underdispersion in the data can be accounted for by specifying an
appropriate model for the abundance parameter, namely, a Bin-CMP model.
More importantly, we allow for large-scale spatial dependence to be
accounted for by adding a spatial component to the intensity model
(i.e., through a spatial basis function expansion). Under the binomial
mixture modeling framework, the use of a low-rank spatial
representation proves to be a computationally advantageous approach to
building in spatial dependence.

Although we have presented a model (\textbf{M2}) that accounts for
covariate information, spatial maps that predict abundance at
unobserved locations could be obtained using model \textbf{M3} and
thereby take advantage of the spline formulation. In contrast, both
models \textbf{M1} and \textbf{M2} would require imputation of
covariates at unobserved locations (i.e., additional data models) to
predict abundance at unobserved locations. Consequently, since our goal
is primarily inferential, this direction has not been pursued here.

The class of binomial mixture models we consider assume population
closure within each primary sampling period. Such an assumption is
often justified based on biological and/or ecological considerations,
when the primary sampling period covers a relatively short time frame.
In our case, the justification of the closed population assumption is
based on ecological considerations. However, it may also be possible to
extend our model to verify the assumption of population closure
following the framework of \citet{dail2011models} by decomposing
the true abundance into the sum of two independent components, that is,
the total number of survivors from the previous sampling period (by
introducing a survival rate parameter in the model) and new additions
prior to the current sampling period (by introducing a birth parameter
in the model). This is a subject of future research.

Although the binomial mixture models we propose can accommodate
unbalanced data structures, the amount of missing data can impact model
selection and parameter estimation. As discussed in the second
simulated example, the model averaged estimates for dispersion
parameters are positively biased when the simulated data exhibit the
same missing pattern as the American Robin data and spatial components
are included to account for spatial dependence in the intensity model.
Nevertheless, we note that grouping of dispersion parameters leads to a
``borrowing of strength,'' since data collected over different years are
pooled together if the corresponding dispersion parameters fall into
the same group. In other words, this pooling of data helps mitigate the
negative impacts of missing values. In general, a comprehensive
assessment of the effect of missing data is problem specific and
depends on both the pattern of missingness and the underlying spatial
dependence (e.g., the effective sample size). In practice, we advocate
evaluating these effects through simulated data examples, similar to
those conducted here.

It is important to note that all of the models we considered for the
American Robin data provide similar results regarding the
identification of important covariates for detectability and intensity,
as well as the grouping of dispersion \mbox{parameters}. First, \texttt
{time,} and \texttt{cloudcover} are identified to be important
covariates for high detectability of the American Robin, with the
former being negatively related to observing the American Robin.
However, one should be careful when interpretating \texttt{cloudcover}
due to the difficulty in estimating it \mbox{objectively} [\citet
{vignola2012solar}]. On the other hand, \texttt{ufbldg}, \texttt
{veg200m,} and \texttt{pubassit} are found to be important predictors
for abundance of the American Robin. In terms of dispersion, the
American Robin data demonstrates overdisperion. Importantly, the class
of binomial mixture models we propose is of \mbox{independent} interest and
when coupled with the CMP distribution can be used in cases where the
type of dispersion (i.e., over- and underdispersion) varies over time.
In this sense, the Bin-CMP mixture model is extremely versatile, as it
can be used for modeling equi-, over-, and underdispersed data (e.g.,
for modeling abundance of less prevalent species, such as the Eastern
Wood Pewee or Wood Thrush in the BES).

\section*{Acknowledgments}
The authors would like to thank the Editor Tilmann Gneiting, Associate Editor, and
three anonymous referees for providing
valuable comments that have helped strengthen this manuscript.

\begin{supplement} [id-suppA]
\stitle{Supplement to ``Bayesian binomial mixture models for estimating abundance in
ecological monitoring studies''}
\slink[doi]{10.1214/14-AOAS801SUPP} 
\sdatatype{.pdf}
\sfilename{aoas801\_supp.pdf}
\sdescription{The supplementary material contains the MCMC sampling algorithm, details regarding computation times for the models implemented, and additional figures.} 
\end{supplement}



%

\printaddresses
\end{document}